\documentstyle[aps]{revtex}
\begin{document}
\draft
\title{Stability of the perturbative vacuum against spatial variations
of the Polyakov loop}
\author{Kenji Fukushima\thanks{E-mail: fuku@nt1.c.u-tokyo.ac.jp}
and Koichi Ohta\thanks{E-mail: ohta@nt1.c.u-tokyo.ac.jp}}
\address{Institute of Physics, University of Tokyo, 3-8-1 Komaba,
Meguro-ku, Tokyo 153-8902, Japan}
\maketitle
\begin{abstract}
We investigate the effective action of the Polyakov loop with spatial
variations. We expand the effective action not in powers of
derivatives or momenta, but in powers of variational amplitudes. At
one-loop order the results suggest that the instability towards the
confining vacuum may be caused by the variational terms.
\end{abstract}

\section{Introduction}
It is believed that the vacuum of the SU($N$) Yang-Mills theories
should be in the colour-confined phase at zero or low temperature,
while at sufficiently high temperature the colour-deconfined phase
would emerge where perturbative calculations do well or at least
somewhat make sense.

In the SU($N$) pure Yang-Mills theories at finite temperature the
centre symmetry is known to be deeply connected with the criteria of
colour confinement\cite{pol78,sve82,sve86}. The order parameter of
the symmetry is given by the expectation value of the Polyakov loop
operator,
\begin{equation}
\Omega=\frac{1}{V}\int{\rm d}^3x\left<\Omega(\bbox{x})\right>=\frac{1}
 {V}\int{\rm d}^3x\left<\frac{1}{N}\text{tr}{\cal T}\exp\left(-{\rm i}
 \int_0^\beta{\rm d}x_4A_4(x_4,\bbox{x})\right)\right>.
\label{eq:ord_par}
\end{equation}
Here $\beta=1/T$ is the inverse temperature, ${\cal T}$ is the time
ordering operator and $\text{tr}$ denotes the trace over the colour
index in the fundamental representation. One can readily relate it to
the free energy $F_q$ in the presence of a single quark with
infinitely heavy mass: $\Omega=\exp(-\beta F_q)$. Therefore, the
confined phase ($F_q=\infty$) corresponds to $\Omega=0$, whereas
$\Omega\neq0$ in the deconfined phase. If the vacuum is centre
symmetric, that is to say $\Omega=L\Omega$ ($L\in{\rm Z}(N)$), it
follows that $\Omega=0$, i.e.\ that the realization of the centre
symmetry is accompanied by the confining vacuum.

The effective potential of the Polyakov loop has been calculated both
in the perturbative approach\cite{gro81,wei81,bel91,bha92,alt94} and
in the strong-coupling expansion on the lattice\cite{pol82,gro83}. The
perturbative results have revealed that the centre symmetry is broken
spontaneously due to the appearance of the Debye mass at high
temperature. It was also proven that any perturbative calculation
inevitably leads to $\Omega=1$, that is the perturbative
vacuum\cite{goc93}, though the proof is based on the suspicious
assumption that the Haar measure could be harmlessly negligible.
Recently, however, Boh\'{a}\v{c}ik\cite{boh90} and Engelhardt and
Reinhardt\cite{eng98} have pointed out in the Polyakov gauge (static
gauge $\partial_4A_4=0$ with an additional condition that $A_4$ should
be diagonal) that the effective action, which includes not only
potential terms but also kinetic terms, has a singularity right at the
perturbative vacuum. The singularity stems from the Haar measure, whose
meaning is still a little confused\cite{bor97,len98,fuk00}. This fact
seems to suggest that the Haar measure or some configurations with
spatial variations such as calorons (multi-instantons at finite
temperature), magnetic monopoles and domain walls might be relevant to
the dynamics of the confinement-deconfinement mechanism.

These former results, however, are somewhat ambiguous with respect to
the two points, i.e.\ {\em renormalization} and {\em gauge invariance}.
Because of the singular character of the Polyakov gauge, it is likely
that gauge spurious objects might arise from incomplete
renormalization. Thus instead of using the singular Polyakov gauge we
prefer the covariant background field gauge in our calculation as in
\cite{bha92,alt94} at the cost of the renormalization necessary for
the Polyakov loop.

One might think that the kinetic energy must be dominated by the tree
term and that the loop corrections should be much smaller than the
tree term so as to make sense of the perturbative calculation.
However, the longitudinal gluon propagator, for instance, acquires the
Debye mass from the contribution of the loop corrections and then in
the soft momentum region the tree term $\sim(\partial A)^2$
proportional to squared momentum can be certainly negligible compared
with the finite Debye mass. Of course, when the momentum becomes
larger the tree term grows to be dominant eventually. Consequently,
one has to take account of loop corrections until the momentum exceeds
the order of $T$. Because we are interested in the momentum region
around $T$, the effective action should be expanded in powers of
variational amplitudes instead of derivatives or
momenta\cite{bha92,eng98}.

The outline of this paper is as follows. We define the effective
action of the Polyakov loop in Sec.\ \ref{sec:def_ef_ac}. Using the
same technique as the derivation of the gauge dependence identity, we
show that the effective action of the Polyakov loop is generally gauge
invariant even if the classical fields do not satisfy the on-shell
conditions (the equations of motion). Sec.\ \ref{sec:per_eva} is
divided into three subsections. Firstly, we expand the one-loop
effective action to quadratic variational terms. The result suggests
possible instability for the perturbative vacuum. Secondly, quartic
terms are found to stabilize the slipped vacuum. Finally we remark
upon the zero temporal modes and the resultant cubic term in the
potential. In Sec.\ \ref{sec:num} the obtained effective action is
examined numerically to give rise to the vacuum instability. We
conclude our discussion and comment upon problems inherent in our
result in Sec.\ \ref{sec:discuss}.

\section{Definition of the effective action}
\label{sec:def_ef_ac}
The effective action in terms of gauge fields $A_\mu^a$ is defined in
a field-theoretic manner by the equation
\begin{equation}
\exp\left(\Gamma[\bar{A}]\right)=\int{\cal D}A{\cal D}\bar{c}{\cal D}c
 \exp\left(S[c,\bar{c},A]+S_{\text{src}}\right)
\end{equation}
where the action $S$ is the sum of the classical action
$S_{\text{cl}}$, the gauge-fixing action $S_{\text{gf}}$ and the ghost
action $S_{\text{ghost}}$, namely,
\begin{eqnarray}
S[c,\bar{c},A]&=&S_{\text{cl}}[A]+S_{\text{gf}}[A]
 +S_{\rm ghost}[c,\bar{c},A] \nonumber\\
&=&\int^\beta{\rm d}^4x\left\{-\frac{1}{4g^2}F_{\mu\nu a}F_{\mu\nu}^a
 -\frac{1}{2g^2\xi}{\cal F}_a{\cal F}^a-\bar{c}_a
 \frac{\delta{\cal F}^a}{\delta A_\mu^b}D_\mu c^b\right\}
\end{eqnarray}
and the source action $S_{\text{src}}$, which only removes the
one-particle reducible diagrams, is written as
\begin{equation}
S_{\text{src}}=\int^\beta{\rm d}^4xJ_{\mu a}(A_\mu^a-\bar{A}_\mu^a)
 \qquad
J_{\mu a}=-\frac{\delta\Gamma[\bar{A}]}{\delta\bar{A}_\mu^a}.
\label{eq:pre_src_act}
\end{equation}
Here $\xi$ is the gauge parameter and ${\cal F}^a$ is to be chosen
appropriately to fix a gauge. The upper and lower indices of colour
should be distinguished, as we employ the ladder basis later (the
Cartan metric in the ladder basis is not diagonal). For the purpose
of defining the effective action of the Polyakov loop, we replace a
part of the degrees of freedom, that is described by the elementary
fields, with the (extended) Polyakov loops defined as
\begin{equation}
\Omega^n(\bbox{x})=\frac{1}{N}\text{tr}\left\{{\cal T}\exp\left(
 -{\rm i}\int_0^\beta{\rm d}x_4A_4(x_4,\bbox{x})\right)\right\}^n
\label{eq:pol_loops}
\end{equation}
where $n$ runs over $1,2,\dots,N-1$ corresponding to the Cartan
subalgebra (the neutral gluons), or alternatively $1,2,\dots,N$ with
a constraint to force the untraced Polyakov loops to be Hermitian.
Because the Polyakov loops are static and written only by the gauge
potential $A_4$, it is natural to transfer the variables from
$\tilde{A}_4^n(\bbox{x})$ to $\Omega^n(\bbox{x})$, where
\begin{equation}
\tilde{A}_4^n(\bbox{x})=\frac{1}{\beta}\int_0^\beta{\rm d}x_4A_4^n
 (x_4,\bbox{x}).
\label{eq:azero}
\end{equation}
Then the source action (\ref{eq:pre_src_act}) can be replaced by
\begin{eqnarray}
S_{\text{src}}&=&\int^\beta{\rm d}^4x\left\{\tilde{J}_{\mu a}
 (A_\mu^a-\bar{A}_\mu^a)+K_n(\Omega^n-\bar{\Omega}^n)\right\}
\label{eq:src_act}\\
\tilde{J}_{\mu a}&=&-\frac{\delta\Gamma[\bar{A},\bar{\Omega}]}
{\delta\bar{A}_\mu^a} \qquad
 \left(\int_0^\beta{\rm d}x_4\tilde{J}_{4n}(x_4,\bbox{x})=0\right)
 \nonumber\\
K_n&=&-\frac{\delta\Gamma[\bar{A},\bar{\Omega}]}
 {\delta\bar{\Omega}^n}
 \nonumber
\end{eqnarray}
where the source $\tilde{J}_{\mu a}$ is almost the same as that in
the source action (\ref{eq:pre_src_act}) except that no more
$\tilde{A}_4^n$ arises out of it. The expression in the brackets
represents this constraint on $\tilde{J}_{\mu a}$. The source for
$\tilde{A}_4^n$ is not $\tilde{J}_{\mu a}$ but $K_n$, in which the
Polyakov loops are regarded as the dynamical variables in place of
$\tilde{A}_4^n$. Then setting $\bar{A}=0$ we define the effective
action of the Polyakov loops that is denoted as
$\Gamma[\bar{\Omega}]$.

In the same manner as the derivation of the gauge dependence
identity\cite{kob91}, it is easy to show that the effective action
defined above is completely independent of a gauge choice of either
${\cal F}$ or $\xi$. The infinitesimal change of the gauge-fixing
condition,
\begin{equation}
{\cal F}^a[A]\rightarrow{\cal F}^a[A]+\delta{\cal F}^a[A] \qquad
\xi\rightarrow\xi+\delta\xi
\label{eq:inf_gauge}
\end{equation}
induces the variation only for the gauge-fixing action as the
following:
\begin{eqnarray}
\delta S_{\text{gf}}[A]&=&-\int^\beta{\rm d}^4x\left(\frac{1}{\xi}
 {\cal F}_a\delta{\cal F}^a-\frac{\delta\xi}{2\xi^2}{\cal F}_a
 {\cal F}^a\right) \nonumber\\
&=&-\int^\beta{\rm d}^4x\frac{1}{\xi}{\cal F}_a\delta'{\cal F}^a
 \nonumber\\
&=&-\int^\beta{\rm d}^4x{\rm d}^4y\frac{\delta S[A]}
 {\delta A_\mu^a(x)}(D_\mu)_{\;b}^a(x){\cal G}_{\;c}^b(x-y)
 \delta'{\cal F}^c(y)
\end{eqnarray}
where ${\cal G}$ is the full propagator of the ghost fields defined
by the equation of motion,
\begin{equation}
\int^\beta{\rm d}^4y\frac{\delta{\cal F}^a(x)}{\delta A_\mu^b(y)}
 (D_\mu)_{\;c}^b(y){\cal G}_{\;d}^c(y-z)=-\delta_{\;d}^a\delta(x-z).
\end{equation}
It is the essential point that $\delta S_{\text{gf}}$ can be
compensated by an appropriate gauge transformation,
\begin{equation}
\delta A_\mu^a(x)=(D_\mu)_{\;b}^a(x)\int^\beta{\rm d}^4y
 {\cal G}_{\;c}^b(x-y)\delta'{\cal F}^c(y).
\label{eq:comp_trans}
\end{equation}
The only terms varied under the subsequent transformations
(\ref{eq:inf_gauge}) and (\ref{eq:comp_trans}) are the source terms,
to say more precisely, fields coupled with the source, provided that
the integration measure contains the factor
$1/\sqrt{\xi}$\cite{kob91}. In the present case the source terms are
given by (\ref{eq:src_act}) among which one term with
$\tilde{J}_{\mu a}$ vanishes because we have set $\bar{A}=0$. In the
other terms with $K_n$, the coupled quantum fields transformed by
(\ref{eq:comp_trans}) are the Polyakov loops $\Omega^n$. They are
definitely invariant for any periodic gauge transformation. Therefore,
it is obvious that the effective action of the Polyakov loops does not
depend on a gauge choice at all due to the gauge invariance of the
Polyakov loops even when the classical fields $\bar{\Omega}^n$ do not
satisfy the stationary condition.

For the purpose of the perturbative calculation it is convenient to
expand the source term (\ref{eq:src_act}) in powers of the elementary
gauge fields. We parameterize the classical Polyakov loops as
\begin{equation}
\bar{\Omega}^n=\frac{1}{N}\text{tr}{\rm e}^{-{\rm i}n\beta a}=
 \frac{1}{N}\sum_{i=1}^N{\rm e}^{-{\rm i}n\beta a_i}
\label{eq:param_Pol}
\end{equation}
with the Hermitian constraint,
\begin{equation}
\sum_{i=1}^Na_i=0.
\label{eq:hermit}
\end{equation}
Later on we will make use of the background field method with the
background fields $a=\text{diag}(a_1,a_2,\dots,a_N)$ introduced in
(\ref{eq:param_Pol}). It is important to note that this definition of
the background fields contains the {\em renormalization} of the
Polyakov loops from the beginning\cite{bel91}. The gauge fields are
decomposed into the quantum parts and the background parts
respectively,
\begin{equation}
A_\mu(x)={\cal A}_\mu(x)+\delta_{\mu4}a(\bbox{x}).
\label{eq:decomp}
\end{equation}
Using equations (\ref{eq:pol_loops}), (\ref{eq:src_act}),
(\ref{eq:param_Pol}) and (\ref{eq:decomp}), we can expand the source
term as follows:
\begin{eqnarray}
&&\int^\beta{\rm d}^4xK_n(\Omega^n-\bar{\Omega}^n) \nonumber\\
&=&-\int^\beta{\rm d}^4x\frac{\delta\Gamma}{\delta\bar{\Omega}^n}
 \frac{1}{N}\text{tr}\left[\left\{{\cal T}{\rm e}^{-{\rm i}
 \int_0^\beta{\rm d}x_4'a}\sum_{l=0}^\infty\frac{1}{l!}\left(-{\rm i}
 \int_0^\beta{\rm d}x_4{\cal A}_4(x_4)\right)^l\right\}^n
 -{\rm e}^{-{\rm i}n\beta a}\right] \nonumber\\
&=&-\int^\beta{\rm d}^4x\frac{\delta\Gamma}{\delta\bar{\Omega}^n}
 \frac{1}{N}\left[-{\rm i}n{\rm e}^{-{\rm i}n\beta a_m}\frac{1}{\beta}
 \int_0^\beta{\rm d}x_4\frac{{\cal A}_4^m(x_4)}{\sqrt{2}}\right.
\nonumber\\
&&-\frac{n\beta}{2}{\rm e}^{-{\rm i}n\beta a_i}\frac{1}{\beta}
 \int_0^\beta{\rm d}x_4\int_0^{x_4}{\rm d}x_4'
 {\rm e}^{{\rm i}(x_4-x_4')(a_i-a_j)}{\cal A}_4^{(i,j)}(x_4)
 {\cal A}_{4(i,j)}(x_4') \nonumber\\
&&\left.-\frac{1}{2}\sum_{k=1}^{n-1}k{\rm e}^{-{\rm i}n\beta a_j
 -{\rm i}k\beta(a_i-a_j)}\int_0^\beta{\rm d}x_4\int_0^\beta{\rm d}x_4'
 {\rm e}^{{\rm i}(x_4-x_4')(a_i-a_j)}{\cal A}_4^{(i,j)}(x_4)
 {\cal A}_{4(i,j)}(x_4')\right]+\text{O}({\cal A}^3) \nonumber\\
&=&-\int^\beta{\rm d}^4x\frac{\delta\Gamma}{\delta a_n}
 \tilde{{\cal A}}_4^n+\int^\beta{\rm d}^4x\frac{1}{2}\left(\frac{
 \delta\Gamma}{\delta a_n}-\frac{\delta\Gamma}{\delta a_m}\right)
 \frac{1}{1-{\rm e}^{-{\rm i}\beta(a_m-a_n)}} \nonumber\\
&&\times\frac{{\rm i}}{\beta}\int_0^\beta{\rm d}x_4\int_0^{x_4}{\rm d}
 x_4'{\rm e}^{{\rm i}(x_4-x_4')(a_n-a_m)}{\cal A}_4^{(m,n)}(x_4)
 {\cal A}_{4(m,n)}(x_4')+\text{O}({\cal A}^3)
\label{eq:src_expan}
\end{eqnarray}
with the notation ${\cal A}_{4(i,j)}$ for the $(i,j)$ component of
the matrix ${\cal A}_4$ divided by $\sqrt{2}$, that is, an
off-diagonal $(i,j)$ component of the ladder basis. In the last line
$\tilde{{\cal A}}_4^n$ is defined by (\ref{eq:azero}) in which
${\cal A}$ is substituted for $A$. If we proceed to the $n$-loop
calculation of the effective action $\Gamma$, we have to use the
$(n-1)$-loop result for $\Gamma$ into $\delta\Gamma/\delta a$ and at
the same time the neglected terms $\text{O}({\cal A}^3)$ are also
necessary.

In the final expression of (\ref{eq:src_expan}) the first term simply
removes all the one-particle reducible diagrams as in the case for the
elementary gauge fields written in (\ref{eq:pre_src_act}). The second
term corresponds to the renormalization of the Polyakov loop, which is
{\em not} the renormalization of the ultraviolet divergence and is
necessary for the two-loop or higher-order calculations of the
effective potential\cite{bel91,alt94}. We emphasize that any effect of
the renormalization of the Polyakov loop automatically comes out in
our formalism because we began with the appropriate definition of the
effective action. Of course, it is also possible to understand the
second term from the point of view of \cite{bel91}. The background
field $a$ is renormalized to $a+\delta a$ and then the effective
action can be expanded as
$\Gamma[a+\delta a]=\Gamma[a]+\delta\Gamma/\delta a\cdot\delta a$, the
second term of which is the same form as the second term in
(\ref{eq:src_expan}). In the present case of the effective
{\em action} the situation is slightly involved. The zero-loop order
of the kinetic term (tree term) may survive so that the effective
action of the one-loop order can also be affected by the
renormalization of the Polyakov loop with spatial variations. However,
although finite corrections may be certainly generated, they can be
absorbed into the arbitrariness of the renormalization of ultraviolet
divergence and thus we do not have to bother with the renormalization
of the Polyakov loop up to one-loop order at all. In the next section
we explore this point further.

\section{Perturbative evaluation of the effective action}
\label{sec:per_eva}
The background field method has an advantage in that the effective
action preserves the manifest gauge invariance\cite{abb81}. The gauge
fields are decomposed into
\begin{equation}
A_\mu^a={\cal A}_\mu^a+A_{\text{B}\mu}^a
\end{equation}
where ${\cal A}_\mu^a$ and $A_{\text{B}\mu}^a$ denote the quantum and
the background part, respectively. Introducing the covariant
derivative in terms of the background fields by
$D_{\text{B}\mu}=\partial_\mu+{\rm i}[A_{\text{B}\mu},\;]$, one can
write the gauge-fixing condition as
\begin{equation}
D_{\text{B}\mu}{\cal A}_\mu^a-\omega^a=0.
\end{equation}
Multiplying a Gaussian weight and integrating $\omega^a$ out, we
obtain the Lagrangian,
\begin{eqnarray}
&&{\cal L}=-\frac{1}{4g^2}\left[F_{\text{B}\mu\nu a}
 F_{\text{B}\mu\nu}^a+4F_{\text{B}\mu\nu a}D_{\text{B}\mu}
 {\cal A}_\nu^a\right. \nonumber\\
&&\qquad-2{\cal A}_{\mu a}\left\{\left(D_{\text{B}}^2\right)_{\;b}^a
 \delta_{\mu\nu}-\left(1-\frac{1}{\xi}\right)(D_{\text{B}\mu}
 D_{\text{B}\nu})_{\;b}^a\right\}{\cal A}_\nu^b \nonumber\\
&&\qquad\left.-4f_{bc}^{\;\:a}D_{\text{B}\mu}{\cal A}_{\nu a}
 {\cal A}_\mu^b{\cal A}_\nu^c+f_{ab}^{\;\:c}f_{dec}{\cal A}_\mu^a
 {\cal A}_\nu^b{\cal A}_\mu^d{\cal A}_\nu^e\right]+\bar{c}_a\left\{
 \left(D_{\text{B}}^2\right)_{\;b}^a+(D_{\text{B}\mu})_{\;c}^a
 f_{bd}^{\;\:c}{\cal A}_\mu^d\right\}c^b.
\end{eqnarray}
From the quadratic part of this Lagrangian, we can readily integrate
the quantum fields ${\cal A}_\mu^a$ and $c^a$ in the Feynman gauge
$\xi=1$ to give the one-loop effective action,
\begin{equation}
\Gamma[A_{\text{B}}]=\int^\beta{\rm d}^4x\left(-\frac{1}{4g^2}
 F_{\text{B}\mu\nu a}F_{\text{B}\mu\nu}^a\right)-\frac{1}{2}\text{tr}'
 \ln\left\{\left(D_{\text{B}}^2\right)_{ab}\delta_{\mu\nu}
 +2f_{ab}^{\;\:c}F_{\text{B}\mu\nu c}\right\}+\text{tr}\ln
 \left\{D_{\text{B}}^2\right\}
\label{eq:eff_act}
\end{equation}
where $\text{tr}$ is the trace of the spatial and colour indices,
while $\text{tr}'$ includes the trace over the Lorentz indices in
addition. We set the background field as (\ref{eq:decomp}) and $a$ is
decomposed into
\begin{equation}
a(\bbox{x})=\frac{1}{\beta}(C+\epsilon(\bbox{x}))
\end{equation}
where $C$ and $\epsilon(\bbox{x})$ represent the constant and the
variational part of the background field, respectively. They are
allowed to be traceless independently, i.e.
\begin{equation}
\text{tr}C=\sum_{i=1}^NC_i=0 \qquad \text{tr}\epsilon(\bbox{x})=
 \sum_{i=1}^N\epsilon_i(\bbox{x})=0.
\end{equation}
We divided them by $\beta$ in order to make $C$ and
$\epsilon(\bbox{x})$ dimensionless. In the ladder
basis\cite{bha92,alt94} the effective action (\ref{eq:eff_act}) is
attained most simply as the following:
\begin{eqnarray}
&&\Gamma[C,\epsilon]=\frac{V}{\beta^3Ng^2}\sum_{i>j}\epsilon_{ij}
 \nabla^2\epsilon_{ij}-\frac{1}{2}\text{tr}'\ln\left[
 \partial_{(i,j)}^2\delta_{\mu\nu}+2{\rm i}\left\{(\partial_\mu
 \delta_{\nu4}-\delta_{\mu4}\partial_\nu)\epsilon_{ij}\right\}
 \right. \nonumber\\
&&\qquad\qquad\left.+2{\rm i}\epsilon_{ij}\delta_{\mu\nu}
 \partial_{4(i,j)}-(\epsilon_{ij})^2\delta_{\mu\nu}\right]+\text{tr}
 \ln\left\{\partial_{(i,j)}^2+2{\rm i}\epsilon_{ij}\partial_{4(i,j)}-
 (\epsilon_{ij})^2\right\}
\label{eq:eff_act2}
\end{eqnarray}
where we rescaled all the coordinate variables by $\beta$ so as to let
them be dimensionless. From now on, we consider all the coordinates,
momenta and fields as dimensionless multiplied appropriately by
$\beta$. Here we denote the dimensionless derivatives with the
background fields as
\begin{eqnarray}
&&\partial_{\mu(i,j)}=\partial_{\mu}+{\rm i}\delta_{\mu4}C_{ij}\qquad
 p_{\mu(i,j)}=p_\mu+\delta_{\mu4}C_{ij}\\
&&C_{ij}=C_{i}-C_{j}\qquad\qquad\epsilon_{ij}=\epsilon_i-\epsilon_j.
\end{eqnarray}
Our purpose is to expand the effective action (\ref{eq:eff_act2}) in
powers of $\epsilon$. In terms of Feynman's diagrammatical method, the
terms with $\epsilon$ and $\epsilon^2$ in the action
(\ref{eq:eff_act2}) are interpreted as the three- and four-point
interaction vertices. The quadratic derivative terms correspond to the
free propagator.

The one-loop effective potential shown diagrammatically in
Fig.\ \ref{fig:zero-var} has already been argued in \cite{wei81} with
the result,
\begin{eqnarray}
\Gamma^{(0)}[C]&=&-\frac{V}{2\beta^3}\int\frac{{\rm d}^3p}{(2\pi)^3}
 \sum_n\text{tr}'\ln\left\{p_{(i,j)}^2\delta_{\mu\nu}\right\}+
 \frac{V}{\beta^3}\int\frac{{\rm d}^3p}{(2\pi)^3}\sum_n\text{tr}\ln
 \left\{p_{(i,j)}^2\right\} \nonumber\\
&=&-\frac{\pi^2V}{12\beta^3}\sum_{i>j}\left(\frac{C_{ij}}{\pi}
 \right)_{\text{mod2}}^2\left\{\left(\frac{C_{ij}}{\pi}
 \right)_{\text{mod2}}-2\right\}^2
\label{eq:wei_poten}
\end{eqnarray}
where $n$ in the summation is the Matsubara frequency, which appears
only through the momentum variables
$p_{(i,j)}=(2\pi n+C_{ij},\bbox{p})$.

\subsection{Up to quadratic order}
The first correction up to $\epsilon^2$ order is represented
diagrammatically in Fig.\ \ref{fig:2-var}. Introducing the Fourier
transform of the variational fields,
\begin{equation}
\epsilon_{ij}(\bbox{x})=\int\frac{{\rm d}^3q}{(2\pi)^3}\epsilon_{ij}
 (\bbox{q}){\rm e}^{-{\rm i}\bbox{q}\cdot\bbox{x}}
\end{equation}
we can easily write the tree term as
\begin{equation}
\Gamma_{\text{tree}}[\epsilon]=-\frac{V}{\beta^3Ng^2}\sum_{i>j}\int
 \frac{{\rm d}^3q}{(2\pi)^3}|\epsilon_{ij}(\bbox{q})|^2\bbox{q}^2.
\end{equation}

The procedure to calculate the loop corrections is almost the same as
that in \cite{bha92} except for the absence of the condition that the
momentum is sufficiently small. This condition is really the case for
the Z($N$) domain wall but in contrast with it our results should
contain all orders of the derivative expansion so that it would not
take the form of the ordinary kinetic term proportional to momentum
squared. The renormalized contribution to the effective action is
calculated as
\begin{eqnarray}
\Gamma^{(2)}_{\text{loop}}[C,\epsilon]&=&\frac{V}{\beta^3}\sum_{i,j}
 \int\frac{{\rm d}^3q}{(2\pi)^3}|\epsilon_{ij}(\bbox{q})|^2\sum_n\int
 \frac{{\rm d}^3p}{(2\pi)^3}\left\{\frac{2(p_{4(i,j)}^2+q^2)}
 {(p_{(i,j)}+q)^2}-1\right\}\frac{1}{p_{(i,j)}^2} \nonumber\\
&=&-\frac{V}{\beta^3}\sum_{i>j}\int\frac{{\rm d}^3q}{(2\pi)^3}
 |\epsilon_{ij}(\bbox{q})|^2\left\{\frac{11}{48\pi^2}q^2\ln\frac{q^2}
 {M^2}+\frac{1}{4\pi}f(q,C_{ij})\right\}
\label{eq:pre_quad_act}
\end{eqnarray}
where we used the 4-vector notation $q=(0,\bbox{q})$. $M$ is a
renormalization point. The first term in brackets in
(\ref{eq:pre_quad_act}) comes from the vacuum part and the second one
comes from the matter part which vanishes as $T\rightarrow0$.

The function $f(q,C)$ is defined by
\begin{eqnarray}
f(q,C)&=&\frac{2}{\pi}\int_0^\infty{\rm d}p\left\{p+q\left(1-
 \frac{p^2}{q^2}\right)\ln\left|\frac{2p-q}{2p+q}\right|\right\}
 N(p,C_{ij}),\\
N(p,C_{ij})&=&\frac{1}{{\rm e}^{p+{\rm i}C_{ij}}-1}
 +\frac{1}{{\rm e}^{p-{\rm i}C_{ij}}-1},
\end{eqnarray}
whose behaviour is shown in Fig.\ \ref{fig:f}. $N(p,C)$ is the
distribution function which is reduced to the usual Bose distribution
function at $C=0$ and to the negative Fermi distribution function at
$C=\pi$. This exchange of the statistical character agrees with the
centre-symmetric treatment proposed recently in which the charged
gluon must obey the antiperiodic boundary condition\cite{len98} (of
course, the origin is actually identical). If the coefficient of
$N(p,C)$ is positive definite and also sufficiently large, the
minimum of the effective action will shift into $C\sim\pi$, i.e.\  the
confining vacuum. In our formalism the result of \cite{boh90,eng98}
can be interpreted as that non-zero value of
$|\epsilon_{ij}(\bbox{q})|$ causes $C_{ij}\rightarrow\pi$ as the
temperature is lowered. However, as we will discuss in the subsequent
sections, our consequence suggests another scenario: $C_{ij}$ remains
zero and only $|\epsilon_{ij}(\bbox{q})|$ makes the Polyakov loop
decrease. Furthermore, the behaviour in the vicinity of $C=\pi$ will
prove to be so infrared singular for spatial variations that it is out
of reach of the perturbative regime.

We have to explain now that the renormalization of the Polyakov loop
is actually harmless to the above result, as mentioned before. The
second term of (\ref{eq:src_expan}) corresponds to the first
correction of the Taylor expansion in the variation from the
renormalization of the Polyakov loop, that is, $\delta\epsilon$. It is
possible to calculate $\delta\epsilon$ from (\ref{eq:src_expan}) at
the lowest order of $\epsilon$ to convince ourselves that
$\delta\epsilon$ is independent of $C_{ij}$ so that the correction to
the effective action takes the form of (constant)$\times q^2
|\epsilon_{ij}(\bbox{q})|^2$ for the quadratic order of $\epsilon$.
This can be absorbed by the redefinition of $M$. Consequently, we can
safely ignore any effect of the renormalization of the Polyakov loop
up to one-loop order.

The above result can be compacted into
\begin{equation}
\Gamma^{(2)}[C,\epsilon]=-\frac{V}{\beta^3}\sum_{i>j}\int
 \frac{{\rm d}^3q}{(2\pi)^3}|\epsilon_{ij}(\bbox{q})|^2\left\{
 \frac{q^2}{Ng^2(q,t)}+\frac{1}{4\pi}f(q,C_{ij})\right\}
\label{eq:quad_act}
\end{equation}
using the running coupling constant,
\begin{equation}
g^2(q,t)=\frac{g^2}{1+(11Ng^2/48\pi^2)\ln(q^2/M^2)}=\frac{48\pi^2}
 {11N\ln(q^2/\beta^2\Lambda^2_0)}=\frac{24\pi^2}{11N\ln(qt)}
\label{eq:run_coup}
\end{equation}
where $t$ denotes the dimensionless temperature in units of
$\Lambda_0$, the QCD scale parameter.

Asymptotic values of $f(q,C)$ are easily obtained as
\begin{eqnarray}
f(q\rightarrow0,C)&=&\frac{4\pi}{3}-4C+\frac{2C^2}{\pi}
\label{eq:f_to_0}\\
f(q\rightarrow\infty,C)&=&-\frac{3}{2}f(q\rightarrow0,C)=
 -2\pi+6C-\frac{3C^2}{\pi}
\label{eq:f_to_inf}
\end{eqnarray}
It is obvious that (\ref{eq:f_to_0}) at $C=0$ provides the correct
perturbative result of the Debye mass which stabilizes the
perturbative vacuum. The behaviour of (\ref{eq:f_to_inf}) that
$f(q,C)$ turns to be negative as $q$ becomes larger is crucial in our
argument. The perturbative vacuum ($C_{ij}=\epsilon_{ij}(\bbox{q})=0$)
may become unstable supposing that the decreasing rate of $f(q,C)$
overcomes the increasing rate of the tree term (quadratic in
momentum). Actually, it is the case when the coupling constant is
beyond some critical value.

Both $C_{ij}$ and $\epsilon_{ij}$ are determined by the condition that
$-\Gamma[C,\epsilon]$ (the free energy in thermodynamics) should be
minimized. From (\ref{eq:quad_act}) it follows that the vacuum
stability for a given $\epsilon(\bbox{q})$ around the perturbative
vacuum depends on the sign of
\begin{equation}
h(q,t)=\frac{1}{4\pi}\left\{\frac{q^2}{N\alpha(q,t)}+f(q,C=0)\right\}
\end{equation}
with $\alpha(q,t)=g^2(q,t)/4\pi$. Of course $\epsilon_{ij}$ is not
independent of each other under the constraint (\ref{eq:hermit}). At
$C=0$, however, the coefficient function $h(q,t)$ contains no colour
dependence ($ij$ dependence) so that a negative $h(q,t)$ is sufficient
for the occurrence of the vacuum instability. The behaviour of $h(q,t)$
in the case of SU(3) ($N=3$) as the temperature $t$ varies is shown in
Fig.\ \ref{fig:h}.

Perturbative calculations are not reliable at all for $\alpha(q,t)>1$.
Looking at Fig.\ \ref{fig:h} we realize that the momentum region with
$\alpha>1$ indicated by the broken curves grows when we lower the
temperature and at the same time the possible unstable modes in which
$h(q,t)<0$ come to spread. Therefore, whether our perturbative
evaluation is reliable or not near the unstable modes depends on
whether the spreading rate of the unstable modes overcomes the growing
rate of the unreliable momentum region. In fact, Fig.\ \ref{fig:h}
indicates that near the critical temperature where unstable modes
begin appearing, the relevant modes are in the momentum region for
$\alpha<1$. To be more precise, at the critical temperature
$T_{\text{c}}=2.77\Lambda_0$, the unstable mode is $q_{\text{c}}=1.33$
and the corresponding critical coupling constant is
$\alpha_{\text{c}}=1.31/N$. In the case of SU(3), for example,
$\alpha_{\text{c}}=0.44$ which is surely smaller than one.

However, it is subtle to decide from the critical value of the
coupling constant whether or not the perturbative calculation works
well at that point. Nevertheless it is most likely that this
instability actually occurs because we can comprehend it in an
intuitive manner as follows. Under the effect of the temperature, real
(on-shell) particles are excited thermally. The average distance
between the thermal particles is of order of $1/T$ and their
distribution is incoherent. Owing to this incoherency the vacuum tends
to be stable against spatial variations with small momentum or
equivalently with larger-scale variations compared with the average
distance between thermal particles. For variations with momentum
larger than $\sim T$, on the other hand, the vacuum tends toward
instability as the overlap of the vacuum configuration with the
distribution of thermal particles increases. If the strength of the
interaction, that is to say, the tendency to overlap the configuration
is large enough, the vacuum may become unstable.

Once unstable modes emerge, the variational fields $\epsilon_{ij}$ go
on increasing to minimize the free energy unless the higher terms are
taken into account. In order to investigate the property of the
instability more closely we need higher-order terms of
$\epsilon_{ij}$. This is the subject dealt with in the next
subsection.

\subsection{Up to quartic order}
Corresponding diagrams to be evaluated are displayed in
Fig.\ \ref{fig:4-var}. The result obtained after tedious calculations
has singularities at $C=0$, whereas the quartic contributions are
ultraviolet and infrared finite as long as $C\neq0$. The reason for
the presence of singularities is quite obvious: since we expanded the
effective action in powers of $\epsilon/(C{\text{\ or\ }}T)$, our
expansion is not valid at $C=0$ for the temporal zero-mode for which
$T$ is absent. The centrifugal barrier at $C=0$ observed in
\cite{boh90,eng98} seems to arise from these spurious singularities.
Therefore, we must manage infrared divergence by resumming zero modes,
as in the case of ring resummation for infrared divergence. We remark
upon this problem in the next subsection.

In this subsection we evaluate the variational term of quartic order
from which all the temporal zero-mode terms are subtracted to be
resummed afterwards. Once we get rid of the zero mode it can be
expected that the result will be almost independent of $q$ in the
interested region $q\sim1.3$, because the expansion in terms of $q$
becomes a power series of
\begin{equation}
\frac{2\bbox{p}\cdot\bbox{q}}{(2\pi n)^2+\bbox{p}^2}\quad{\text{or}}
\quad\frac{\bbox{q}^2}{(2\pi n)^2+\bbox{p}^2}
\end{equation}
where $\bbox{p}$ is the momentum to be integrated and $n$ is the
Matsubara frequency. Although in the case of $n=0$ they are singular
near $\bbox{p}\sim0$, we can regard them as small enough as far as
$n\neq0$ and $q\sim1.3$. Of course, it is possible to compute the
quartic correction directly from the corresponding diagrams. We
confirmed numerically that the result, which is too intricate an
expression to write down here, hardly depends on $q$ around the
unstable region.

Now that the result proved to be almost $q$-independent we can
evaluate the quartic correction at the limit of $q\rightarrow0$.
\begin{eqnarray}
\Gamma^{(4)}[C,\epsilon]&\rightarrow&-\frac{V}{\beta^3}\sum_{ij}\int
 \frac{{\rm d}^3q_1}{(2\pi)^3}\frac{{\rm d}^3q_{2}}{(2\pi)^3}
 \frac{{\rm d}^3q_3}{(2\pi)^3}\epsilon_{ij}(\bbox{q}_1)
 \epsilon_{ij}(\bbox{q}_2)\epsilon_{ij}(\bbox{q}_3-\bbox{q}_1)
 \epsilon_{ij}(-\bbox{q}_2-\bbox{q}_3) \nonumber\\
&&\times\int\frac{{\rm d}^3p}{(2\pi)^3}\sum_{n\neq0}\left[
 \frac{4p_{4(i,j)}^4}{p_{(i,j)}^8}+\frac{4p_{4(i,j)}^2}{p_{(i,j)}^6}
 +\frac{1}{2p_{(i,j)}^4}\right] \nonumber\\
&=&-\frac{V}{\beta^3}\sum_{i>j}\int\frac{{\rm d}^3q_1}{(2\pi^3)}
 \cdots\int\frac{{\rm d}^3p}{(2\pi)^3} \nonumber\\
&&\times\int_{-\infty+{\rm i}\eta}^{\infty+{\rm i}\eta}
 \frac{{\rm d}p_4}{(2\pi)}\left[\frac{8p_4^4}{p^8}+\frac{8p_4^2}{p^6}
 +\frac{1}{p^4}\right]\left(1+\frac{1}{{\rm e}^{-{\rm i}p_4+{\rm i}
 C_{ij}}-1}+\frac{1}{{\rm e}^{-{\rm i}p_4-{\rm i}C_{ij}}-1}\right)
 \nonumber\\
&=&-\frac{V}{\beta^3}\sum_{i>j}\int\frac{{\rm d}^3q_1}{(2\pi)^3}
 \cdots\int\frac{{\rm d}^3p}{(2\pi)^3}\frac{1}{12}\frac{\partial^3}
 {\partial p^3}\left(1+\frac{1}{{\rm e}^{p+{\rm i}C_{ij}}-1}+\frac{1}
 {{\rm e}^{p-{\rm i}C_{ij}}-1}\right) \nonumber\\
&=&-\frac{V}{\beta^3}\sum_{i>j}\int\frac{{\rm d}^3q_1}{(2\pi)^3}
 \frac{{\rm d}^3q_2}{(2\pi)^3}\frac{{\rm d}^3q_3}{(2\pi)^3}
 \epsilon_{ij}(\bbox{q}_1)\epsilon_{ij}(\bbox{q}_2)
 \epsilon_{ij}(\bbox{q}_3-\bbox{q}_1)
 \epsilon_{ij}(-\bbox{q}_2-\bbox{q}_3)\frac{1}{12\pi^2}.
\label{eq:quart_coef}
\end{eqnarray}
The first, second and third term in the square bracket corresponds to
Fig.\ \ref{fig:4-var}\ ($a$)$-$($c$), respectively. The result
(\ref{eq:quart_coef}) is nothing but the coefficient of $C^4$ in the
Weiss potential (\ref{eq:wei_poten}). Since the Weiss potential is a
quartic polynomial, we can realize without any calculation that terms
of order higher than quartic are never relevant under the soft
momentum limit if subtracted by their zero mode.

\subsection{Zero-mode resummation}
One may wonder why we did not take account of the cubic terms of
$\epsilon$ in spite of the fact that the Weiss potential
(\ref{eq:wei_poten}) has the cubic term $C^3$. If we compute them
perturbatively as in the same manner for quadratic or quartic order,
however, they are always zero: the integrand for the cubic order is
proportional to $p_{4(i,j)}$ so that the summation over the Matsubara
frequency and the colour indices leads to zero. Then where does the
cubic term in the potential (\ref{eq:wei_poten}) come from? The answer
is the zero-mode resummation, which is quite similar to the
$e^3$-correction of the QED partition function\cite{kup89}.

Here we will perform the zero-mode resummation only for the most
dangerous terms that are divergent in the infrared limit. The total
momentum insertion from the variational fields $\epsilon$ is
necessarily zero for one blob (self-energy part) so that each
propagator from one blob to the other possesses the same momentum to
bring about the strongest infrared singularities, as is represented in
Fig.\ \ref{fig:ring_diag}.

Taking away all the terms with $n\neq0$ and subtracting quadratic
terms (infrared finite terms), we reduce the effective action
(\ref{eq:eff_act2}) into
\begin{eqnarray}
&&-\frac{1}{4}\text{tr}'\ln\left[\delta_{\mu\nu}-
 \frac{2\epsilon_{ij}^2}{\partial_{(i,j)}^2}\delta_{\mu\nu}-
 \frac{4C_{ij}^2\epsilon_{ij}}{\partial_{(i,j)}^2}\frac{\epsilon_{ij}}
 {\partial_{(i,j)}^2}\delta_{\mu\nu}+\frac{4\{(\partial_\mu
 \delta_{\lambda4}-\delta_{\mu4}\partial_\lambda)\epsilon_{ij}\}}
 {\partial_{(i,j)}^2}\frac{\{(\partial_\lambda\delta_{\nu4}-
 \delta_{\lambda4}\partial_\nu)\epsilon_{ij}\}}{\partial_{(i,j)}^2}
 \right. \nonumber\\
&&\qquad\left.+\frac{4{\rm i}C_{ij}\epsilon_{ij}}{\partial_{(i,j)}^2}
 \frac{\{(\partial_\mu\delta_{\nu4}-\delta_{\mu4}\partial_\nu)
 \epsilon_{ij}\}}{\partial_{(i,j)}^2}+\frac{4{\rm i}C_{ij}\{(
 \partial_\mu\delta_{\nu4}-\delta_{\mu4}\partial_\nu)\epsilon_{ij}\}}
 {\partial_{(i,j)}^2}\frac{\epsilon_{ij}}{\partial_{(i,j)}^2}+
 \frac{\epsilon_{ij}^2}{\partial_{(i,j)}^2}\frac{\epsilon_{ij}^2}
 {\partial_{(i,j)}^2}\delta_{\mu\nu}\right] \nonumber\\
&&\qquad+\frac{1}{2}\text{tr}\ln\left[1-\frac{2\epsilon_{ij}^2}
 {\partial_{(i,j)}^2}-\frac{4C_{ij}^2\epsilon_{ij}}
 {\partial_{(i,j)}^2}\frac{\epsilon_{ij}}{\partial_{(i,j)}^2}+
 \frac{\epsilon_{ij}^2}{\partial_{(i,j)}^2}\frac{\epsilon_{ij}^2}
 {\partial_{(i,j)}^2}\right]
 -(\epsilon_{ij}^2{\text{-terms}})
\label{eq:n0}
\end{eqnarray}
where any term of $\epsilon^3$ is absent because all the diagrams
possessing such vertices are to be resummed individually to contribute
either to the order of $\epsilon^5$ or to less singular terms. In
order to evaluate (\ref{eq:n0}) we perform calculations in the
momentum space. $\epsilon(\bbox{x})$ is sum of all the modes
$\epsilon(\bbox{q})$, among which we leave only the most singular
combination, as shown in Fig.\ \ref{fig:ring_diag}.

To take the trace for the Lorentz indices, it is useful to decompose
them into three projective components, namely
$P_{{\text{T}}\mu\nu}=\delta_{\mu\nu}-q_\mu q_\nu/q^2-\delta_{4\mu}
\delta_{4\nu}$, $P_{{\text{L}}\mu\nu}=q_\mu q_\nu/q^2$ and
$u_{\mu\nu}=\delta_{4\mu}\delta_{4\nu}$. Then (\ref{eq:n0}) can be
rewritten as
\begin{eqnarray}
&&-\frac{V}{4\beta^3}\sum_{i,j}\int\frac{{\rm d}^3p}{(2\pi)^3}
 \sum_{\mu=\nu}\ln\left[\delta_{\mu\nu}+\left\{\int\frac{{\rm d}^3q}
 {(2\pi)^3}|\epsilon_{ij}(\bbox{q})|^2\left(\frac{2}{p_{(ij)}^2}-
 \frac{4C_{ij}^2}{p_{(i,j)}^2(p_{(i,j)}+q)^2}\right)\right.\right.
 \nonumber\\
&&\qquad\left.+\frac{1}{p_{(i,j)}^4}\left(\int\frac{{\rm d}^3q}
 {(2\pi)^3}|\epsilon_{ij}(\bbox{q})|^2\right)^2\right\}
 P_{{\text{T}}\mu\nu}+\left\{\int\frac{{\rm d}^3q}{(2\pi)^3}
 |\epsilon_{ij}(\bbox{q})|^2\left(\frac{2}{p_{(i,j)}^2}-
 \frac{4(C_{ij}^2+q^2)}{p_{(i,j)}^2(p_{(i,j)}+q)^2}\right)\right.
 \nonumber\\
&&\qquad\left.\left.+\frac{1}{p_{(i,j)}^4}\left(\int\frac{{\rm d}^3q}
 {(2\pi)^3}|\epsilon_{ij}(\bbox{q})|^2\right)^2\right\}(
 P_{{\text{L}}\mu\nu}+u_{\mu\nu})\right] \nonumber\\
&&\qquad+\frac{V}{2\beta^3}\sum_{i,j}\int\frac{{\rm d}^3p}{(2\pi)^3}
 \sum_{\mu=\nu}\ln\left[1+\int\frac{{\rm d}^3q}{(2\pi)^3}
 |\epsilon_{ij}(\bbox{q})|^2\left(\frac{2}{p_{(i,j)}^2}-
 \frac{4C_{ij}^2}{p_{(i,j)}^2(p_{(i,j)}+q)^2}\right)\right.
 \nonumber\\
&&\qquad\left.+\frac{1}{p_{(i,j)}^4}\left(\int\frac{{\rm d}^3q}
 {(2\pi)^3}|\epsilon_{ij}(\bbox{q})|^2\right)^2\right]-
 (\epsilon_{ij}^2{\text{-terms}}) \nonumber\\
&=&-\frac{V}{2\beta^3}\sum_{i,j}\int\frac{{\rm d}^3p}{(2\pi)^3}\left[
 \ln\left\{1+\int\frac{{\rm d}^3q}{(2\pi)^3}
 |\epsilon_{ij}(\bbox{q})|^2\left(\frac{2}{p_{(i,j)}^2}-
 \frac{4(C_{ij}^2+q^2)}{p_{(i,j)}^2(p_{(i,j)}+q)^2}\right)\right.
 \right. \nonumber\\
&&\qquad\left.\left.+\frac{1}{p_{(i,j)}^4}\left(\int\frac{{\rm d}^3q}
 {(2\pi)^3}|\epsilon_{ij}(\bbox{q})|^2\right)^2\right\}-\int
 \frac{{\rm d}^3q}{(2\pi)^3}|\epsilon_{ij}(\bbox{q})|^2\left(
 \frac{2}{p_{(i,j)}^2}-\frac{4(C_{ij}^2+q^2)}{p_{(i,j)}^2
 (p_{(i,j)}+q)^2}\right)\right].
\label{eq:pre_nzero}
\end{eqnarray}
We restrict ourselves to considering the case of $C_{ij}=0$ because no
infrared singularity emerges otherwise. For an isolated mode
$\bbox{q}$, the zero-mode action (\ref{eq:pre_nzero}) is simplified as
\begin{equation}
-\frac{V}{2\beta^3}\sum_{i,j}\int\frac{{\rm d}^3p}{(2\pi)^3}\left[
 \ln\left\{1+|\epsilon_{ij}|^2\left(\frac{2}{p^2}-\frac{4q^2}
 {p^2(p+q)^2}\right)+\frac{|\epsilon_{ij}|^4}{p^4}\right\}-
 |\epsilon_{ij}|^2\left(\frac{2}{p^2}-\frac{4q^2}{p^2(p+q)^2}
 \right)\right].
\label{eq:nzero}
\end{equation}
It is easy to reproduce the cubic term in the Weiss potential
(\ref{eq:wei_poten}) with $\bbox{q}=0$. In this limiting case
(\ref{eq:nzero}) is reduced to
\begin{eqnarray}
&=&-\frac{V}{\beta^3}\sum_{i,j}\frac{1}{2\pi^2}\int_0^\infty{\rm d}p
 \left\{p^2\ln\left(1+\frac{|\epsilon_{ij}|^2}{p^2}\right)-
 |\epsilon_{ij}|^2\right\} \nonumber\\
&=&\frac{V}{\beta^3}\sum_{i,j}\frac{1}{3\pi^2}\int_0^\infty{\rm d}p
 \frac{|\epsilon_{ij}|^4}{p^2+|\epsilon_{ij}|^2}=\frac{V}{\beta^3}
 \sum_{i>j}\frac{|\epsilon_{ij}|^3}{3\pi}.
\end{eqnarray}
This is certainly identical to the cubic term in the Weiss potential.

If the relevant mode $\bbox{q}$ is non-zero then the situation changes
drastically. For sufficiently small values of $|\epsilon_{ij}|\ll q$
the main contribution of the integration over $p$ stems from
singularities near $p\sim0\sim|\epsilon_{ij}|$. Then a similar
calculation as above leads to
\begin{eqnarray}
&\simeq&-\frac{V}{\beta^3}\sum_{i,j}\frac{1}{2\pi^2}\int_0^\infty
 {\rm d}p\left\{p^2\ln\left(1-\frac{|\epsilon_{ij}|^2}{p^2}\right)+
 |\epsilon_{ij}|^2\right\} \nonumber\\
&=&\frac{V}{\beta^3}\sum_{i,j}\frac{1}{3\pi^2}\int_0^\infty{\rm d}p
 \frac{|\epsilon_{ij}|^4}{p^2-|\epsilon_{ij}|^2}=0.
\label{eq:qnzero}
\end{eqnarray}
The integration in the last line is defined in the sense of Cauchy's
principal value. The result (\ref{eq:qnzero}) is, in fact, what is
expected: the $\epsilon^3$-term is generated by the most severe
infrared singularities. Once $q$ takes a non-vanishing value, infrared
singularities within the blobs (not the propagators, see
Fig. \ref{fig:ring_diag}) are loosened.

The result obtained up to now is valid only for
$|\epsilon_{ij}|\ll q$. In the next section, we want to deal with not
so small $\epsilon$ around $q_{\text{c}}$, beyond which the
perturbative evaluation will lose its  validity as shown in
Fig.\ \ref{fig:num_order}. If the contribution from the zero-mode
resummation comes to be opposite to that from the
quartic term for some marginal value of $\epsilon$, the zero-mode
resummation may cause qualitative changes: the absolute minimum of our
effective action may jump with a finite gap. In fact, however, the
zero-mode resummation rapidly grows as $q$ increases from zero so
that it becomes positive contribution (the same sign as the quartic
term) in the momentum region of interest. No qualitative picture can
be affected by the zero-mode resummation. Thus for the simplicity we
will entirely ignore this contribution in the next section. Although
this neglection seems too rough, closer investigation of the zero-mode
resummation reveals that it is not so crude. Numerical cancellation
between the resummed contribution performed here and the
next-to-leading contribution occurs.

The expansion singularities of $\epsilon/C$ resummed in this subsection
have much to do with the famous problem of Linde's infrared
catastrophe\cite{lin80}. Since Fig.\ \ref{fig:2-var} is the same as
the diagram for the gluon vacuum polarization, the two- or higher-loop
calculation should break down and non-perturbative effects are no
longer under control perturbatively. The infrared singularities would
be regulated and perturbative expansion would work well only provided
that the constant background field $C$ is of order of $g^0$ or at
least $g^1$ as is discussed in \cite{alt94}. It is the reason why the
hard thermal loops (HTLs) are not necessary for the evaluation of the
effective potential. The naive order counting by the coupling
constant will work due to the presence of the background. Since the
stationary value of the background is zero, in fact, there seems to be
a logical contradiction in this argument. In our formalism the
variational background field $\epsilon$ instead of $C$ takes a
non-zero value, as seen below. Linde's problem is regulated here into
the expansion singularities of $\epsilon/C$ that can be resummed in
principle. Therefore, we can consider that our perturbative expansion
of the order of $g$ should make sense (this does not say nothing about
the reliability of the perturbative evaluation) and expect that a
non-perturbative contribution will be suppressed once instability
occurs.

\section{Numerical analysis}
\label{sec:num}
The free energy density obtained above is
\begin{eqnarray}
U[C,\epsilon]&=&-\frac{\beta^3}{V}\left(\Gamma^{(0)}[C]
 +\Gamma^{(2)}[C,\epsilon]+\Gamma^{(4)}[\epsilon]\right) \nonumber\\
&=&\sum_{i>j}\left[\frac{\pi^2}{12}\left(\frac{C_{ij}}{\pi}
 \right)_{\text{mod2}}^2\left\{\left(\frac{C_{ij}}{\pi}
 \right)_{\text{mod2}}-2\right\}^2+\int\frac{{\rm d}^3q}{(2\pi)^2}
 |\epsilon_{ij}(\bbox{q})|^2\frac{1}{4\pi}\left\{\frac{q^2}
 {N\alpha(q,t)}+f(q,C_{ij})\right\}\right. \nonumber\\
&&\left.+\frac{1}{12\pi^2}\int\frac{{\rm d}^3q_1}{(2\pi)^3}
 \frac{{\rm d}^3q_2}{(2\pi)^3}\frac{{\rm d}^3q_3}{(2\pi)^3}
 \epsilon_{ij}(\bbox{q}_1)\epsilon_{ij}(\bbox{q}_2)
 \epsilon_{ij}(\bbox{q}_3-\bbox{q}_1)
 \epsilon_{ij}(-\bbox{q}_2-\bbox{q}_3)\right].
\label{eq:density}
\end{eqnarray}
We can acquire the equations of motion from the stationary conditions
for the free energy, i.e.\
$\delta U/\delta C_{ij}=\delta U/\delta\epsilon_{ij}(\bbox{q})=0$ and,
in principle, solve them self-consistently. It is, however, quite
involved because of the mode couplings induced by the nonlinear term
$\epsilon^4$. To simplify the analysis and investigate the situation
in a comprehensible manner we assume that $\epsilon^2(\bbox{x})$ is
such a function that the variance from the spatial mean value is small
enough to be negligible, that is,
\begin{equation}
\frac{\beta^3}{V}\int{\rm d}^3x\left\{\epsilon_{ij}^2(\bbox{x})-
 \frac{1}{V}\int{\rm d}^3x\epsilon_{ij}^2(\bbox{x})\right\}^2\simeq0.
\label{eq:approx}
\end{equation}
Then the nonlinear term takes the form,
\begin{equation}
\frac{1}{12\pi^2}\left(\int\frac{{\rm d}^3q}{(2\pi)^3}|\epsilon_{ij}
(\bbox{q})|^2\right)^2.
\end{equation}
This assumption is correct provided that $\epsilon(\bbox{x})$ is a
step-like function or has no sharp peak.

\begin{figure}
\caption{Diagrams for the one-loop effective potential. The curly
curves and the broken curves represent the gauge fields and the ghost
fields, respectively.}
\label{fig:zero-var}

\caption{Diagrams up to quadratic order. The doubled curly curves
denote the variational fields.}
\label{fig:2-var}

\caption{The behaviour of the function $f(q,C)$ with various values of
$C$.}
\label{fig:f}

\caption{The behaviour of the function $h(q,t)$ for various
temperature $t$. Broken curves represent the momentum region where
$\alpha>1$.}
\label{fig:h}

\caption{The variational corrections of quartic order.}
\label{fig:4-var}

\caption{The ring diagram : each pair of doubled curly curves has the
opposite momentum so that every single curly curve possesses the same
momentum (inserted momentum is always zero).}
\label{fig:ring_diag}

\caption{The behaviour of the order parameter as a function of the
temperature $t$ for SU(2) and SU(3).}
\label{fig:num_order}

\caption{The free energy density only in terms of $C$: The broken
curves are in the unreliable region $\alpha>1$.}
\label{fig:free_ene}
\end{figure}
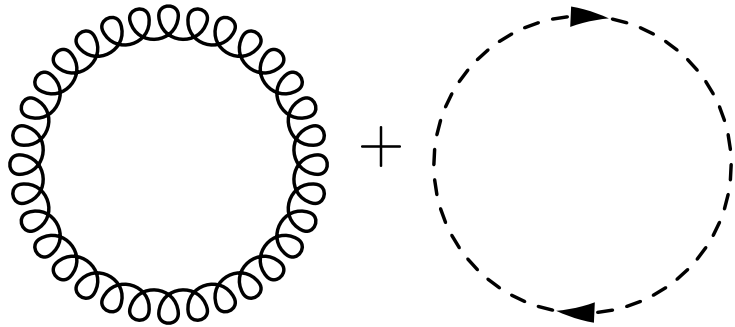
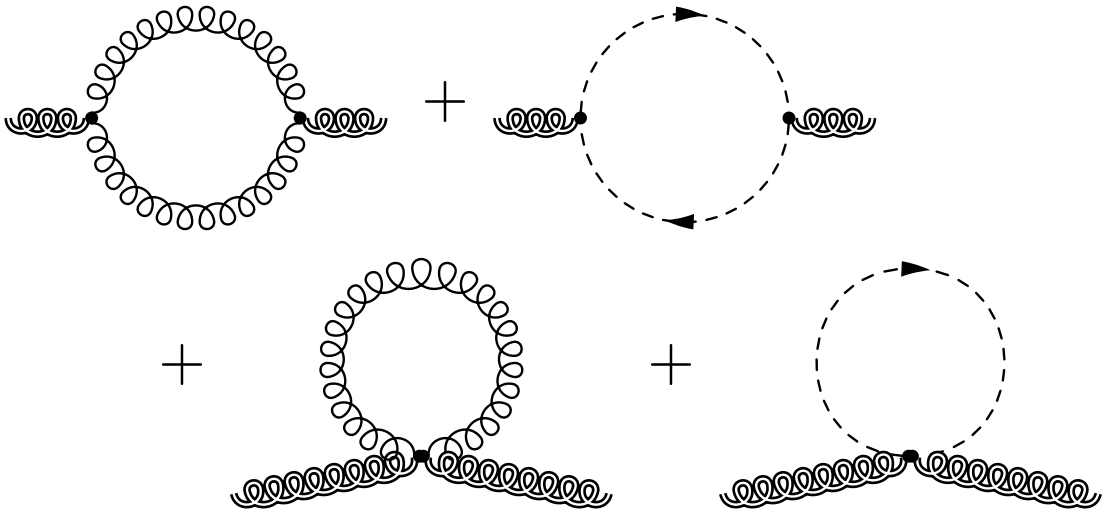
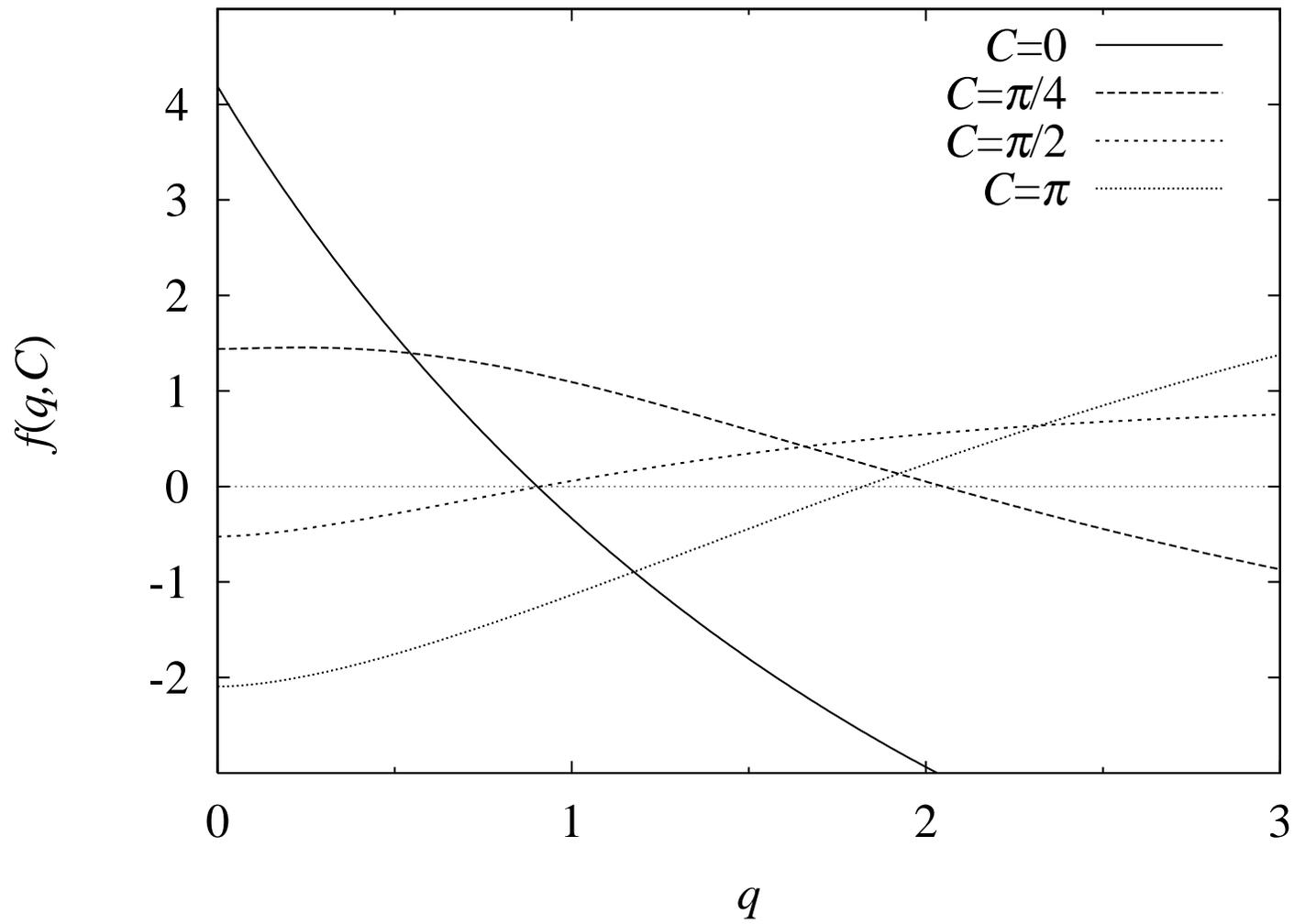
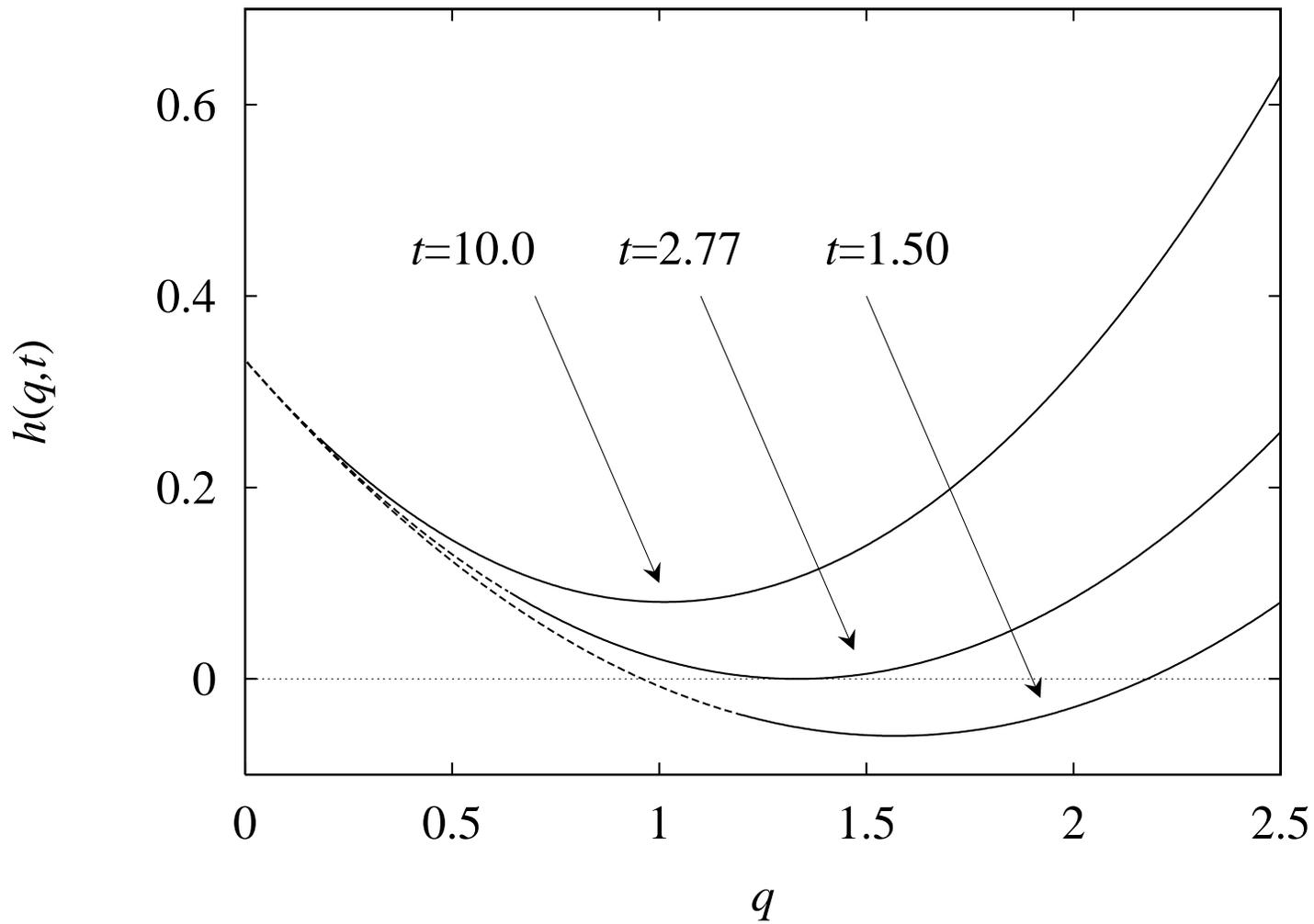
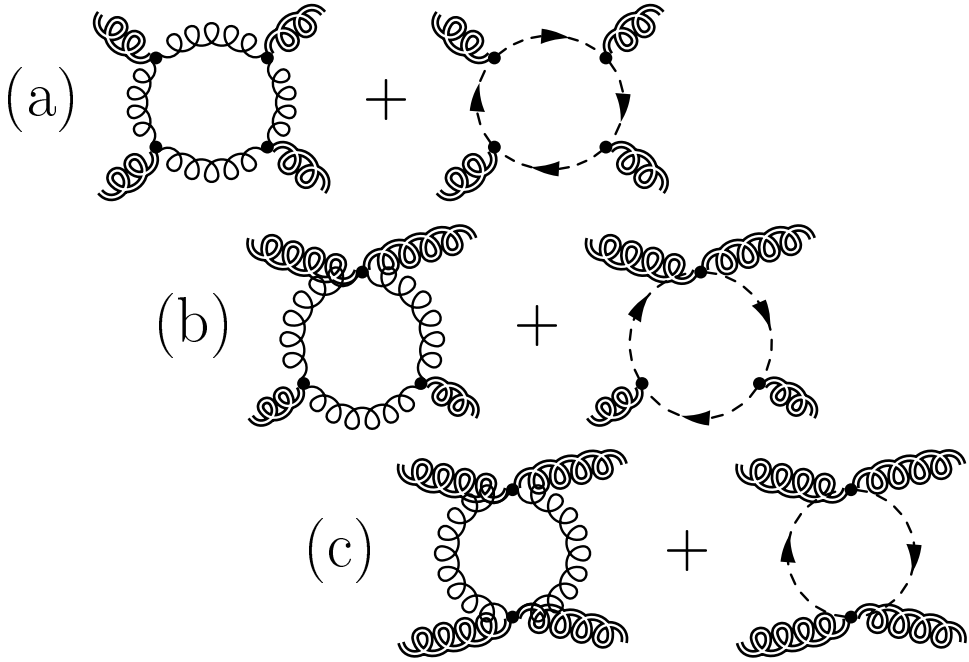
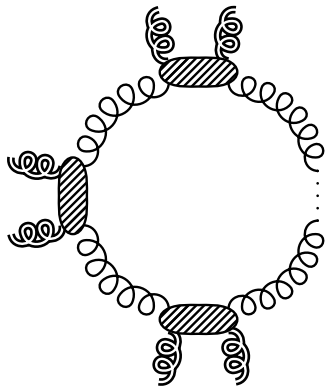
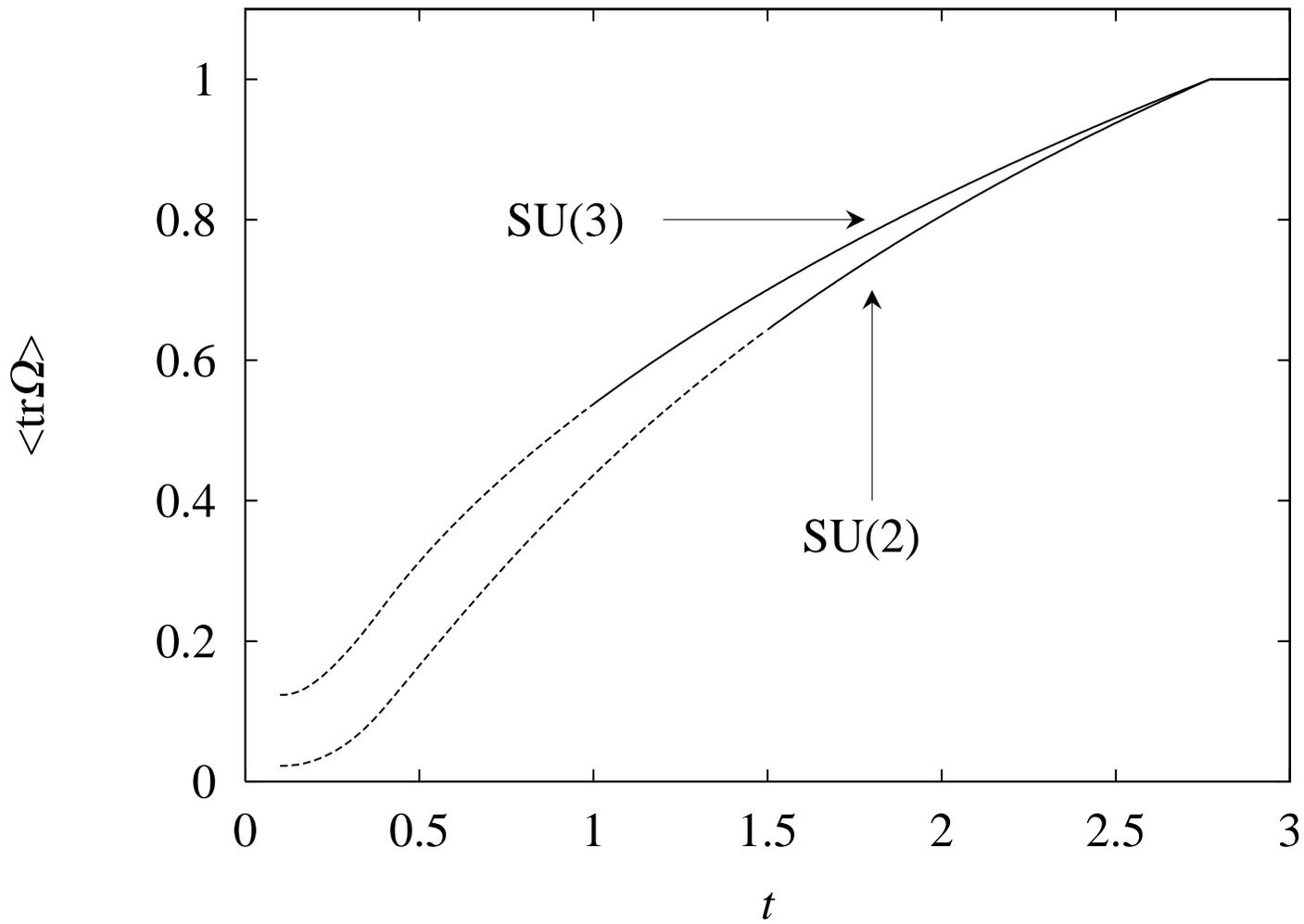
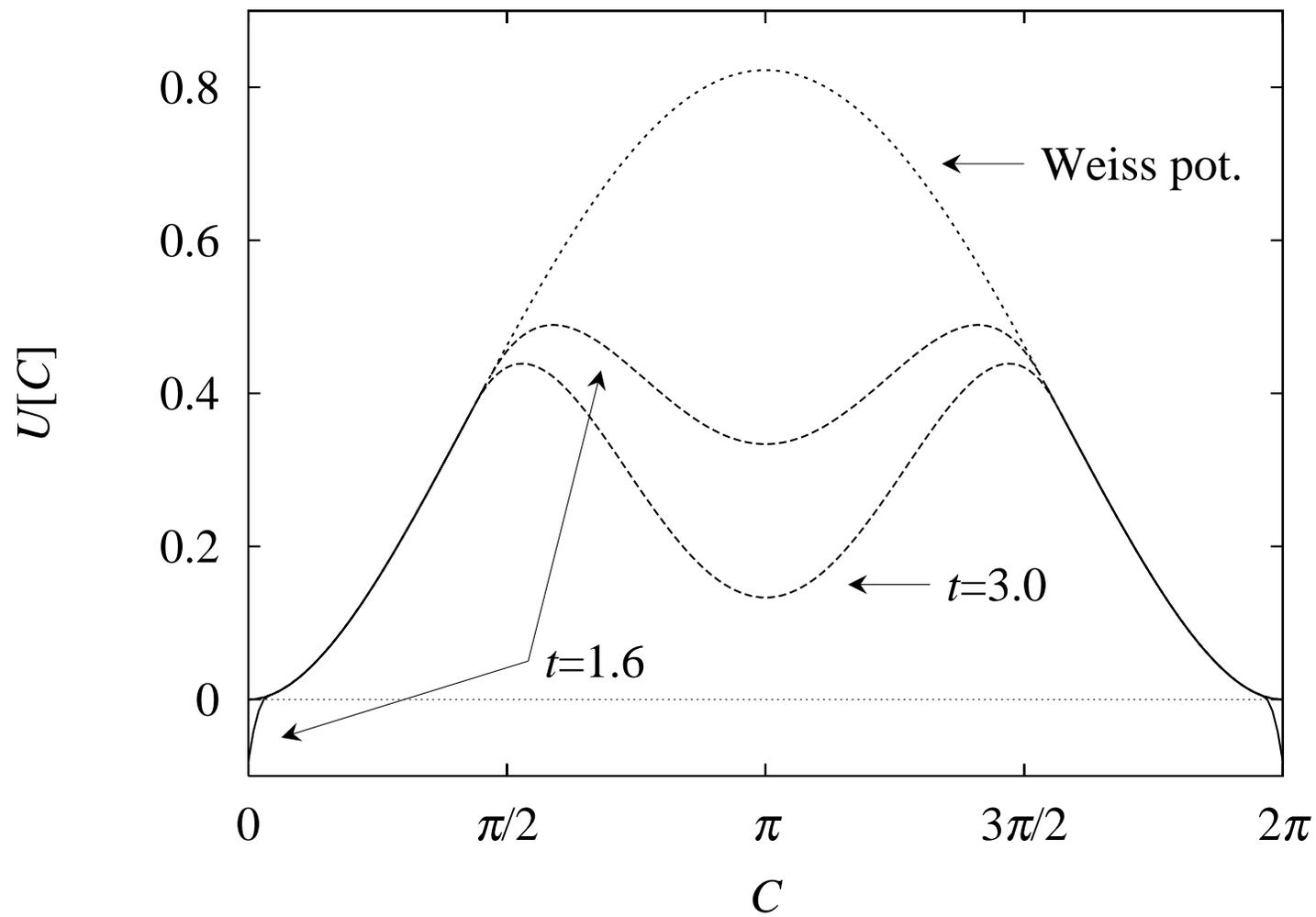

For simplicity we shall initially discuss the SU(2) case where
$\epsilon_2=-\epsilon_1=\epsilon/2$ and $C_2=-C_1=C/2$. From the
equations of motion one can easily obtain
\begin{equation}
\int\frac{{\rm d}^3q}{(2\pi)^3}|\epsilon(\bbox{q})|^2=
 -6\pi^2\tilde{h}(C,t),
\label{eq:su2_em}
\end{equation}
where $\tilde{h}(C,t)$ is defined as
\begin{equation}
\tilde{h}(C,t)=\frac{1}{4\pi}\left\{\frac{q_0^2(C,t)}
 {N\alpha(q_0(C,t),t)}+f(q_0(C,t),C)\right\},
\end{equation}
with $q_0(C,t)$ determined in such a way that
$q^2/N\alpha(q,t)+f(q,C)$ should take its minimum value to minimize
$U[C,\epsilon]$. Then the free energy density is written only in terms
of $C$ as
\begin{equation}
U[C]=U^{(0)}[C]-3\pi^2\{\tilde{h}(C,t)\}^2,
\label{eq:su2_poten}
\end{equation}
where $U^{(0)}[C]$ is the Weiss potential (density),
\begin{equation}
U^{(0)}[C]=\frac{\pi^2}{12}\left(\frac{C}{\pi}\right)_{\text{mod2}}^2
 \left\{\left(\frac{C}{\pi}\right)_{\text{mod2}}-2\right\}^2.
\end{equation}
The order parameter (\ref{eq:ord_par}) takes the following form under
the approximation (\ref{eq:approx}) and using the stationary value
(\ref{eq:su2_em}),
\begin{equation}
\Omega=\exp\left(\frac{3\pi^2}{4}\tilde{h}(C,t)\right)\cos\frac{C}{2}.
\label{eq:su2_order_para}
\end{equation}
Minimizing (\ref{eq:su2_poten}) leads to $C=0$ so that $\Omega$ varies
only through the change of $\epsilon$ determined by
(\ref{eq:su2_em}) at $C=0$. The numerical behaviour of the order
parameter as a function of $t$ is displayed in
Fig.\ \ref{fig:num_order}.

It is straightforward to proceed to the case of SU(3) where the
independent variables are $\epsilon_1$, $\epsilon_2$, $C_1$ and $C_2$.
The traceless constraint (\ref{eq:hermit}) gives
$\epsilon_3=-\epsilon_1-\epsilon_2$ and $C_3=-C_1-C_2$. Numerical
investigation tells us that the minimum lies either at
$\epsilon_1=\epsilon_2$ or at $\epsilon_1=-2\epsilon_2$
($\epsilon_2=-2\epsilon_1$). Whichever of them we may choose, they
are actually equivalent. We take the former case, namely,
$\epsilon_1=\epsilon_2=\epsilon$. The stationary value in this case
corresponding to (\ref{eq:su2_em}) is
\begin{equation}
\int\frac{{\rm d}^3q}{(2\pi)^3}|\epsilon(\bbox{q})|^2=-\frac{\pi^2}{3}
 \left\{\tilde{h}(2C_1+C_2,t)+\tilde{h}(C_1+2C_2,t)\right\},
\end{equation}
and the free energy density corresponding to (\ref{eq:su2_poten})
takes the form of
\begin{eqnarray}
U[C]&=&U^{(0)}[C_2-C_1]+U^{(0)}[2C_1+C_2]+U^{(0)}[C_1+2C_2]\nonumber\\
&&\quad-\frac{3\pi^2}{2}\left\{\tilde{h}(2C_1+C_2,t)
 +\tilde{h}(C_1+2C_2,t)\right\}^2
\label{eq:su3_poten}
\end{eqnarray}
whose minimum turns out to be at $C_1=C_2=0$. The behaviour of the
order parameter shown in Fig.\ \ref{fig:num_order} is almost the same
as in the SU(2) case.

It should be noted that we have discovered the possibility of
instability but we cannot reach the {\em phase transition}. The phase
transition is defined by the critical point above which spontaneous
magnetization (order) appears. Any critical characteristics, such as
the transition temperature, critical exponents and so on, are all
defined at the very critical point. Thus all we can say is that
Fig.\ \ref{fig:num_order} suggests some continuous {\em instability}.
If we want to investigate the critical point, we must start from the
confining vacuum as the strong-coupling limit on the lattice, for
example, or some non-perturbative formalism\cite{fuk00}.

As mentioned above, our result of the free energy density
(\ref{eq:density}) cannot describe any discontinuous transition which
is believed to be the case for SU(3). The reason for this is obvious:
equation (\ref{eq:density}) does not possess any terms of
$\epsilon^3$. Since the centre group Z(3) allows terms of $\Omega^3$
only in the case of SU(3), $\epsilon^3$ terms are expected to be
produced from some origins peculiar to SU(3).

Another possibility to lead to the discontinuity is the instability
near $C\sim\pi$. Even if (\ref{eq:density}) has no odd terms of
$\epsilon$ it is a complicated nonlinear function in terms of $C$ so
that the discontinuous transition in terms of $C$ cannot be ruled out.
Actually, Fig.\ \ref{fig:f} shows that $f(q,C=\pi)$ becomes negative
in the soft momentum region, which may cause $h(q,t)$ to be negative.
Although this instability seems to be similar to the case of $C\sim0$,
the relevant modes are entirely different. On one hand, in the case of
$C\sim0$, the unstable modes are around $q\sim1.3$ that is comparable
to the temperature. In the case of $C\sim\pi$, on the other hand, the
instability is enhanced as the momentum goes to the soft limit
$q\rightarrow0$, where the running coupling constant at the one-loop
order does not make sense. Our result that (\ref{eq:su3_poten}) has
the absolute minimum at $C_1=C_2=0$ would be valid provided that the
minimum value of $\tilde{h}(C=\pi,t)$ were at most
$f(q\rightarrow0,C=\pi)/4\pi=-1/6$ (we have used this assumption
implicitly in searching for the minimum of (\ref{eq:su2_poten}) and
(\ref{eq:su3_poten})). However, the behaviour of $\tilde{h}(C,t)$
obtained here is only up to one-loop order and thus we have no idea
what really happens near $C\sim\pi$. All we can state is that
$C\sim\pi$ is certainly unstable and that the relevant momentum is so
infrared that any perturbative calculations would not do well. It is
therefore probable that the position of the absolute minimum jumps
from $C=0$ to $\pi$ in the SU(3) case if the free energy density is
attained in a non-perturbative way, which is beyond the scope of this
paper.

The situation is clear in Fig.\ \ref{fig:free_ene} which shows the
behaviour of (\ref{eq:su2_poten}). The instability discovered here is
for the modes near $C=0$ at lower temperature. Around $C\simeq\pi$ the
relevant modes are so infrared that non-perturbative effects turn out
to be of importance, even if the temperature is sufficiently high. It
might seem curious that in Fig.\ \ref{fig:free_ene} the potential
value at $t=3.0$ is lower than that at $t=1.6$ near $C=\pi$; this is
because we have searched for the most unstable mode $q_0$ within a
range larger than $1/t$ in order to avoid spurious negative values of
the running coupling constant. When the temperature becomes larger,
the lower bound of the momentum range to be searched decreases and
thus the potential value also decreases.

\section{Summary and conclusion}
\label{sec:discuss}
We investigate the effective action of the Polyakov loop defined in a
field-theoretic manner. The gauge invariance of the effective action
can be proven by the same strategy as deriving the gauge dependence
identities. Owing to the appropriate definition of the effective
action any effect of the renormalization of the Polyakov loop that is
necessary for higher-order calculations is automatically generated
through non-local interaction vertices. In order to evaluate the
effective action we choose the background field gauge where the
background gauge symmetry remains manifestly. In this gauge the most
essential properties to attain reliable results, that is, the gauge
invariance and the unambiguous renormalization, are satisfied
simultaneously.

In the actual calculation we expand the effective action in powers of
the variational amplitudes $\epsilon(\bbox{x})$ instead of the
derivatives. The quadratic term of the variational field shows the
possibility of the vacuum instability for marginal modes around $C=0$
or for the constant background $C\sim\pi$. The reason why the vacuum
may become unstable is comprehensible from our resultant expression:
as for the former case ($C=0$), the variational background is usually
absent because such a configuration costs more kinetic energy. In the
thermal circumstance, however, on-shell particles are excited to obey
the thermal distribution function, in addition to the quantum
excitation. If the mode of the variational background becomes
comparable with the inter-distance of the thermal particles, i.e.\
$\sim1/T$ and if the interaction is sufficiently strong, then the
vacuum can decay into non-vanishing variational background. As for the
latter case ($C\sim\pi$) the statistical character of the transverse
gluon changes into that of the Fermi particle, that is to say, the
transverse gluon obeys the antiperiodic boundary condition. Moreover
the distribution number of particles turns to be negative. The serious
problem we encounter is the magnitude of the coupling constant. It
depends on the smallness of coupling constant whether our effective
action at one-loop order is reliable or not. In the SU(3) case the
critical coupling constant is $\alpha_{\text{c}}=0.44$. Particularly
at finite temperature this value is too large to regard it as small.
It is well known that the higher-order corrections to the free energy
up to $g^5$ order (this is the upper limit of Linde's problem) is so
oscillating that the perturbative series does not seem to converge at
all. Nevertheless we believe that the physical picture may be
reflected qualitatively in the one-loop effective action. It is an
interesting and essential point that the kinetic term of the effective
action even at one-loop order can bring about instability at the
perturbative vacuum, which is out of reach of the conventional
effective potential. The effective action at one-loop order already
has a non-trivial physical content.

Once we pursue the policy of studying the physics contained in the
one-loop effective action or believe in the perturbative evaluation
despite the large coupling value, we can proceed to expand the
effective action to quartic order. A brief discussion reveals that the
quartic term is independent of modes provided that the temporal zero
mode is subtracted. Looking over the Weiss potential we can see that
any terms higher than quartic order come from the temporal zero mode.
The zero-mode resummation is performed, where we make clear the origin
of the cubic term in the Weiss potential.

The effective action density determines the stationary vacuum for the
variational background $\epsilon(\bbox{x})$ and the constant
background $C$. The minimum lies at $C=0$ so that the Polyakov loop
decreases only by non-vanishing $\epsilon$ provided that the
instability near $C=\pi$ is harmless. Since we know that in the SU(2)
case the transition is continuous (second order), the instability
found in this paper is likely to be relevant to the real physics. In
the SU(3) case, on the other hand, the discontinuous transition
implies the possibility that the infrared instability around $C=\pi$
comes to play an essential role.

It is necessary, of course, that the qualitative argument discussed
here is justified by some reliable method other than perturbative
evaluations. Some methods, such as the thermal renormalization group
and infrared effective theory successfully predict the
non-perturbative contribution for the infrared region. The application
of these methods to estimate the effective action is quite difficult
due to the presence of the non-local intricate vertices corresponding
to the renormalization of the Polyakov loop. That is the future work
to resolve.

\subsection*{acknowledgments}
One of the authors (KF) is supported by Research Fellowships of the
Japan Society for the Promotion of Science for Young Scientists.

\end{document}